\documentclass[3p,times]{elsarticle}
\usepackage{ecrc}
\volume{00}
\firstpage{1}
\journalname{Nuclear Physics A}
\runauth{T. Lappi and H. M\"antysaari}
\jid{npa}
\jnltitlelogo{Nuclear Physics A}

\usepackage{mathrsfs}
\usepackage{amssymb}
\usepackage{amsmath}
\usepackage{subdepth}

\usepackage{xcolor}
\definecolor{lcolor}{rgb}{0.5,0,0}
\definecolor{citcolor}{rgb}{0,0.3,0.0}

\usepackage[breaklinks,colorlinks,urlcolor=blue,citecolor=citcolor,linkcolor=lcolor]{hyperref}
\usepackage{wrapfig}

\biboptions{square,comma,numbers,sort&compress}

\usepackage[figuresright]{rotating}

\newcommand{\re}{Ref.~}
\newcommand{\eq}{Eq.~}
\newcommand{\eqs}{Eqs.~}
\newcommand{\nr}[1]{(\ref{#1})}
\newcommand{\res}{Refs.~}

\newcommand{\fig}{Fig.~}

\newcommand{\ud}{\, \mathrm{d}}

\newcommand{\xt}{{\mathbf{x}}}
\newcommand{\yt}{{\mathbf{y}}}
\newcommand{\ut}{{\mathbf{u}}}
\newcommand{\vt}{{\mathbf{v}}}
\newcommand{\zt}{{\mathbf{z}}}

\newcommand{\kt}{{\mathbf{k}}}

\newcommand{\Kt}{{\mathbf{K}}}

\newcommand{\xit}{{\boldsymbol{\xi}}}
\newcommand{\etat}{{\boldsymbol{\eta}}}

\newcommand{\astil}{{ \widetilde{\alpha} }}
\newcommand{\lqcd}{\Lambda_{\mathrm{QCD}}}
\newcommand{\as}{\alpha_{\mathrm{s}}}
\newcommand{\nc}{{N_\mathrm{c}}}
\newcommand{\qs}{Q_\mathrm{s}}

\newcommand{\tr}{\, \mathrm{Tr} \, }

\newcommand{\ktt}{k_T} % scalar
\begin{document}

\begin{frontmatter}



\dochead{}
%% e.g. \dochead{17th International Conference on Dynamical Processes in Excited States of Solids}

\title{Proposal for a running coupling JIMWLK equation}


\author[jyfl,hip]{T. Lappi}
\author[jyfl]{H. M\"antysaari }

\address[jyfl]{Department of Physics, %
 P.O. Box 35, 40014 University of Jyv\"askyl\"a, Finland}
\address[hip]{Helsinki Institute of Physics, P.O. Box 64, 00014 University of Helsinki,
Finland}

\begin{abstract}
In the CGC framework the initial stages of a heavy ion collision at high energy are described as ``glasma'' field configurations. The initial condition for these evolving fields depends, in the CGC effective theory, on a probability distribution for color charges. The energy dependence of this distribution can be calculated from the JIMWLK renormalization group equation.
We discuss recent work~\cite{Lappi:2012vw} on a practical implementation
of the running coupling constant in the Langevin method of solving the JIMWLK equation.

\end{abstract}

\begin{keyword}



\end{keyword}

\end{frontmatter}

% \linenumbers


\section{Introduction}
The Color Glass Condensate
(CGC) framework (for reviews see e.g.~\cite{Gelis:2010nm,Lappi:2010ek})
provides a powerful effective theory for QCD at the high collision 
 energies  reached at the LHC.
In the CGC picture, the most convenient choice for the basic degrees of freedom
of a hadronic target are Wilson lines (path ordered exponentials of the 
color field), which describe the eikonal propagation of a high energy 
probe through the target. In the CGC, these are stochastic variables drawn from 
an energy-dependent probability distribution
that satisfies the  JIMWLK  renormalization group
equation. Different scattering processes are related to 
expectation values of various operators 
formed from the Wilson lines. 

In phenomenological work the simplest operator, the two point function or
``dipole'',  can most conveniently be obtained from
the Balitsky-Kovchegov (BK)
equation~\cite{Balitsky:1995ub,Kovchegov:1999yj,Kovchegov:1999ua}
which, although the original derivation predates JIMWLK, can now be viewed
as its mean field approximation.
The inclusion of the running coupling constant in the BK 
equation (rcBK)~\cite{Kovchegov:2006vj,Balitsky:2006wa,Albacete:2007yr} has been
essential  for successful phenomenological 
applications~\cite{Albacete:2009fh,Albacete:2010sy,Albacete:2010bs,Kuokkanen:2011je,Albacete:2012xq,Lappi:2012nh,Lappi:2013zma}.
The running of the QCD coupling slows down the energy dependence 
of cross sections bringing it to rough  agreement
with HERA measurements. Since the full NLO version of the 
equation~\cite{Balitsky:2008zza} has not been extensively applied to phenomenology
(see however recent work in \re\cite{Avsar:2011ds}), rcBK is currently 
the state of the art in phenomenological applications.

The few existing solutions~\cite{Lappi:2011ju,Dumitru:2011vk}
of the JIMWLK equation with a running coupling have
 not used the same ``Balitsky''~\cite{Balitsky:2006wa} prescription commonly used with the 
BK equation. This is due to the need
to decompose the evolution kernel into a product of two factors
for the numerical solution of the JIMWLK equation.
We describe here a recent proposal~\cite{Lappi:2012vw}
for a form of the running coupling JIMWLK equation, given by 
\eqs\nr{eq:jtimestepsymmeta} and~\nr{eq:newnoisecorr}, and argue 
that it results in parametrically the same scale for the coupling constant
as in the  ``Balitsky'' prescription for the BK equation.
Our starting point is different from 
\res\cite{Kovchegov:2006vj,Balitsky:2006wa,Albacete:2007yr}, where 
one sets out to find the correct modification of the BK kernel, keeping the form
of the BK equation otherwise intact.
Here we in stead keep the functional form of the
JIMWLK equation intact, with the result that the
mean field BK approximation does not have the conventional form.

\section{The JIMWLK equation}

The  equation for the probability distribution of Wilson lines
can be expressed as a
functional Langevin equation for the Wilson lines
themselves~\cite{Blaizot:2002xy},
\begin{equation}\label{eq:langevin1}
\frac{\ud}{\ud y} V_\xt = V_\xt (i t^a) \left[
\int_\zt 
\varepsilon_{\xt,\zt}^{ab,i} \; \xi_\zt(y)^b_i  + \sigma_\xt^a 
\right] .
\end{equation}
Here we denote two dimensional vectors by $\xt,\cdots$.  The coordinate arguments are
written  as subscripts. The Wilson line $V$ is a unitary matrix in the
fundamental
representation, generated by $t^a$  and $i=1,2$ is a transverse spatial index.
The first term in \eq\nr{eq:langevin1} is proportional to a stochastic noise $\xi$
and the second one is a deterministic ``drift'' term.
The coefficient of the noise in the stochastic term is
the ``square root'' of the  JIMWLK Hamiltonian
\begin{equation}
 \varepsilon_{\xt,\zt}^{ab,i} = \left(\frac{\as}{\pi}\right)^{1/2}\;
K_{\xt-\zt}^i
\left[1-U_\xt^\dag  U_\zt\right]^{ab},
\textnormal{ where }
K^i_{\xt-\zt} = \frac{ (\xt-\zt)^i }{ (\xt-\zt)^2}
\end{equation}
is the light cone wave function for soft gluon emission.
The noise $\xit = (\xi_1^a t^a,\xi_2^a t^a ) $
is taken to be Gaussian and local in color, transverse coordinate
and rapidity $y$ (evolution time) with
$\langle \xi_\xt(y)^b_i\rangle =0$ and 
$
\langle \xi_\xt(y)^a_i \xi_\yt(y')^b_j\rangle = \delta^{ab}
\delta^{ij}\delta^{(2)}_{\xt\yt} \delta(y-y').
$
Using the Langevin form one can easily derive the evolution equation for the dipole
$
 \hat{D}_{\xt,\yt}  = 
 \frac{1}{\nc}  \tr V^\dag_\xt V_\yt 
$
as
\begin{equation}\label{eq:fixedas}
\partial_y \hat{D}_{\xt,\yt}(y) 
=
\frac{\as \nc}{2 \pi^2}
\int_{\zt}
\bigg(
\Kt_{\xt-\zt}^2 
+
  \Kt_{\yt-\zt}^2
-2 \Kt_{\xt-\zt}\cdot\Kt_{\yt-\zt}
\bigg)
\left[
\hat{D}(\xt,\zt)\hat{D}(\zt,\yt) 
- \hat{D}(\xt,\yt)
\right],
\end{equation}
from which the BK equation is obtained by taking the expectation value and replacing 
$\langle \hat{D}\hat{D}\rangle \to\langle \hat{D}\rangle \langle \hat{D}\rangle $.
It has been known for some time that the JIMWLK equation 
can be written in an explicitly left-right symmetric form~\cite{Kovner:2005jc}.
The corresponding symmetric form for the Langevin equation was first written down 
in~\cite{Lappi:2012vw} as
\begin{equation}\label{eq:jtimestepsymm}
V_\xt(y + \ud y) = 
\exp\left\{-i\frac{\sqrt{\as \ud y }}{\pi}\int_\zt 
  \Kt_{\xt-\zt} \cdot ( V_\zt \xit_\zt V^\dag_\zt) \right\}
V_\xt 
\exp\left\{i\frac{\sqrt{\as \ud y }}{\pi}\int_\zt 
  \Kt_{\xt-\zt} \cdot \xit_\zt \right\},
\end{equation}
and greatly simplifies the numerical solution of the equation
due to the absence of the drift term.

\section{The scale of the coupling}

\begin{figure}
\centerline{
\includegraphics[width=0.3\textwidth]{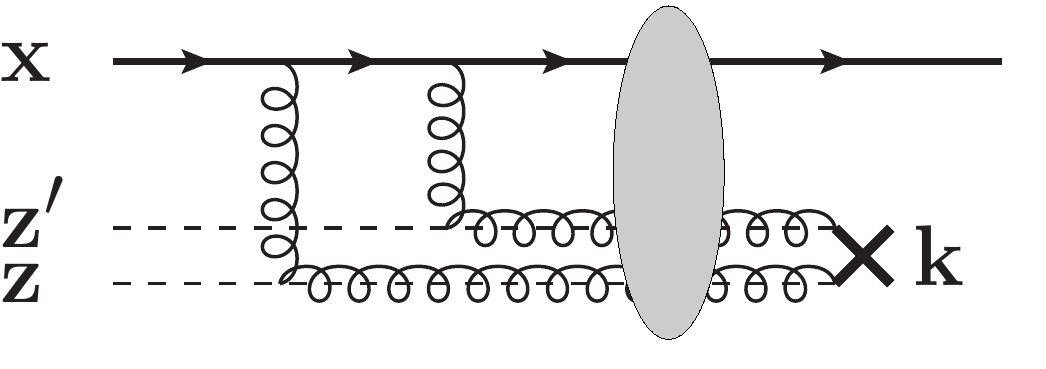}
\includegraphics[width=0.3\textwidth]{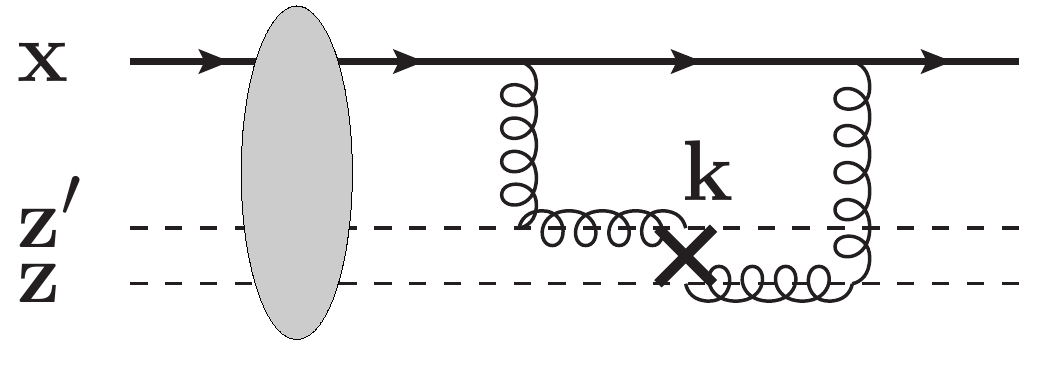}
\includegraphics[width=0.3\textwidth]{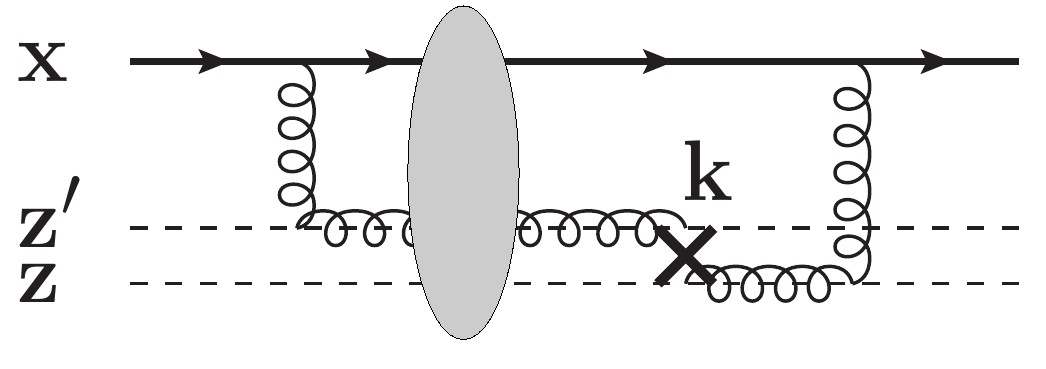}}
\caption{\label{fig:virt}
Virtual diagrams with coupling constant scale depending on the 
momentum of the gluon after the shockwave.
Left: two $\xi$'s from the left of the Wilson line in
 \eq\nr{eq:jtimestepsymmeta} transported through the target (factors $V_\zt$)
and contracted. Center: contraction of
one $\xi$ from the left and one from the right. Right:
contraction of to $\xi$'s on the right.
}
\end{figure}
\begin{figure}
\centerline{
\includegraphics[width=0.3\textwidth]{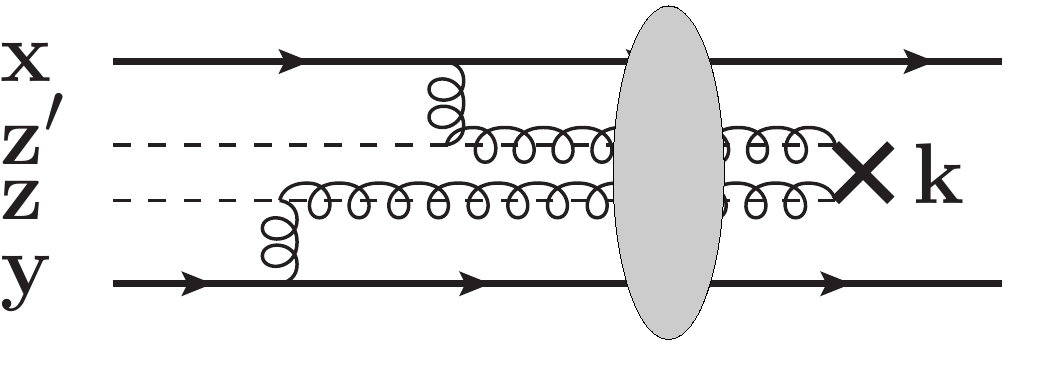}
\includegraphics[width=0.3\textwidth]{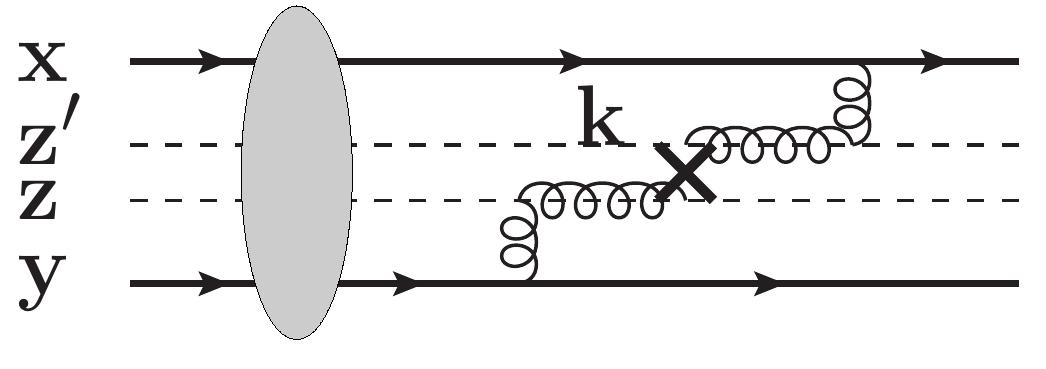}
\includegraphics[width=0.3\textwidth]{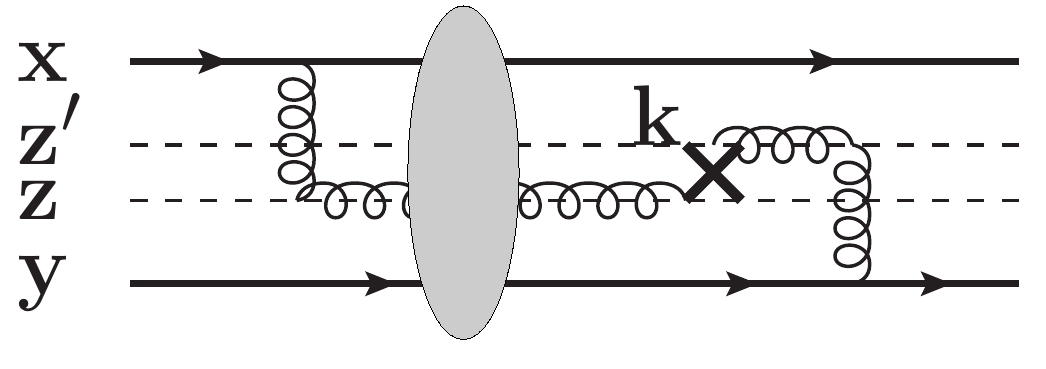}}
\caption{\label{fig:real}
Real diagrams with coupling constant scale depending on the 
momentum of the gluon after the shockwave.
}
\end{figure}

The rapidity evolution of  a general correlator of $n$ Wilson lines 
is obtained by changing all the Wilson lines
by \eq\nr{eq:jtimestepsymm}, developing to order $\xit^2$ (i.e. to order
$\ud y$) and taking the expectation values with the probability distribution 
of the noise $\xit$. Physically keeping only the quadratic order in $\xit$ 
corresponds to the fact that JIMWLK is a LO evolution equation, derived
by considering the emission of only one gluon. The interpretation
of the contractions between two $\xit$'s is that one corresponds
to the emission of a gluon in the amplitude and the other one 
in the complex conjugate.
In the formulation that includes the deterministic term this 
interpretation is not as straightforward, since in \eq\nr{eq:langevin1}
one is summing a noise term corresponding to a gluon emission amplitude
and the deterministic term which is an emission probability, i.e. amplitude squared.

\begin{figure}
\centerline{
\includegraphics[width=0.45\textwidth]{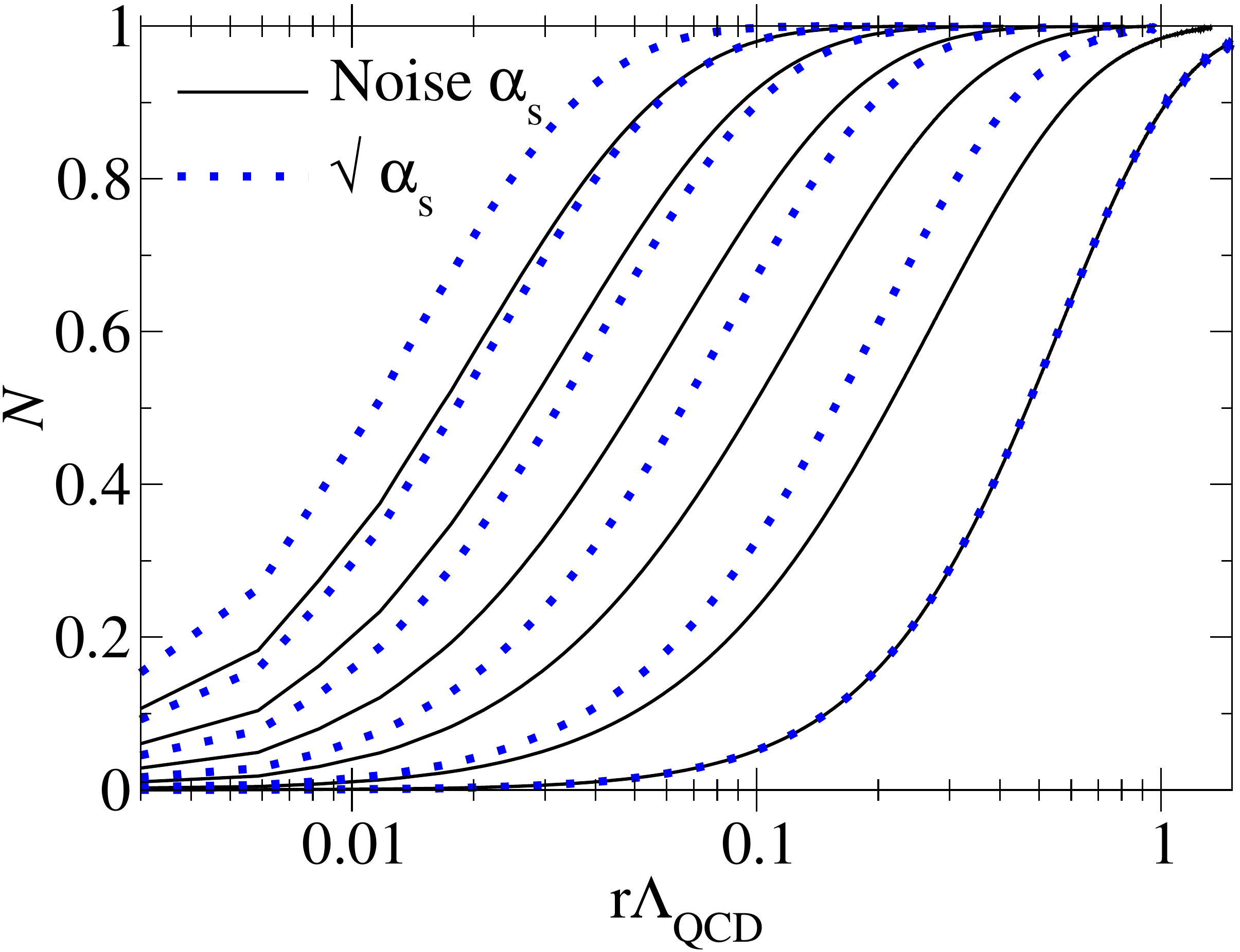}
\includegraphics[width=0.45\textwidth]{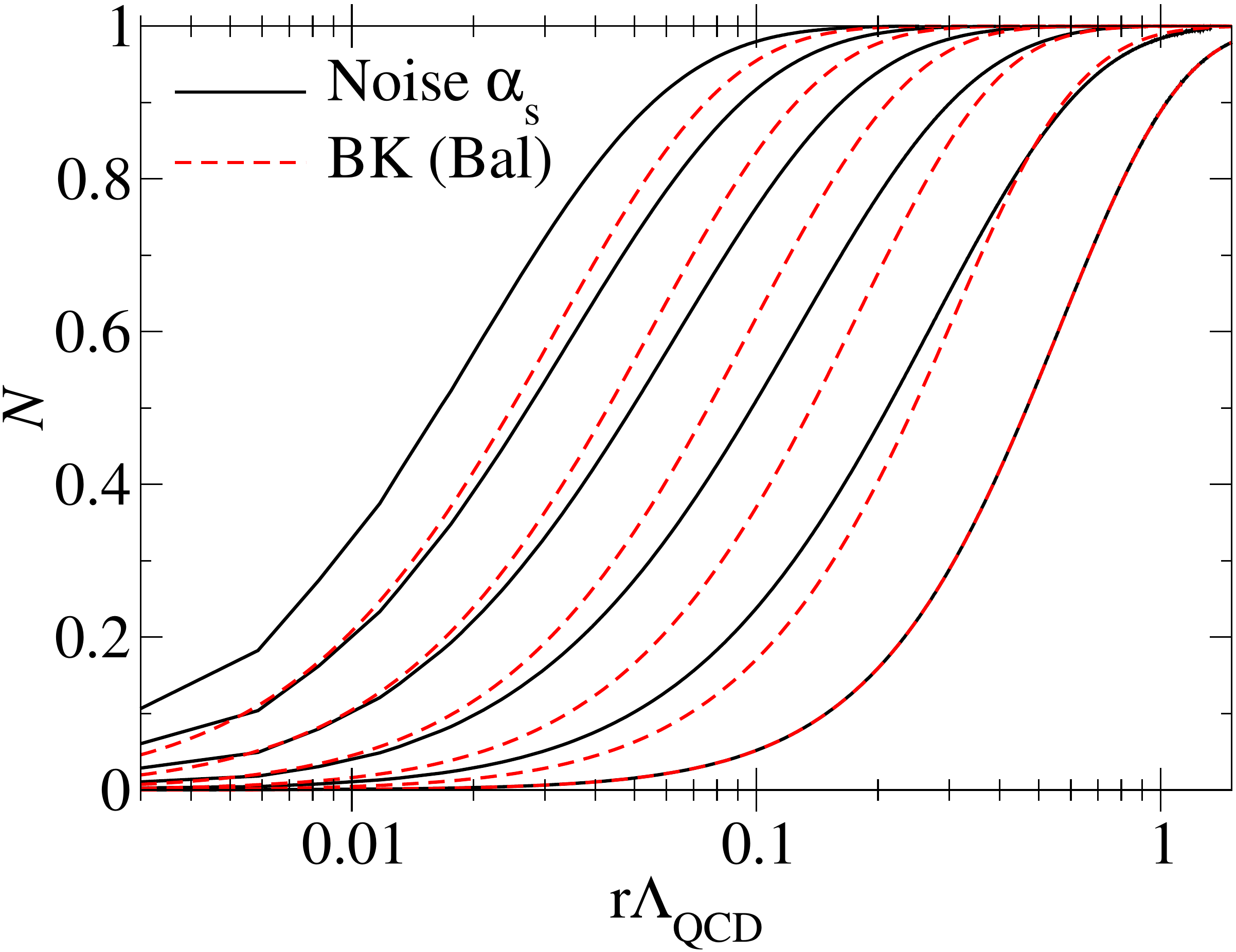}}
\caption{\label{fig:dipoles}
Left: the dipole operator evolved with \eq\nr{eq:jtimestepsymmeta} compared to the 
``square root'' prescription used in~\cite{Lappi:2011ju,Dumitru:2011vk}. Both evolutions
are started from the same initial condition (curves on the right).
Right: comparison between \eq\nr{eq:jtimestepsymmeta} and the BK equation with the 
``Balitsky'' prescription.
}
\end{figure}

\begin{wrapfigure}{r}{0.5\textwidth}
\includegraphics[width=0.45\textwidth]{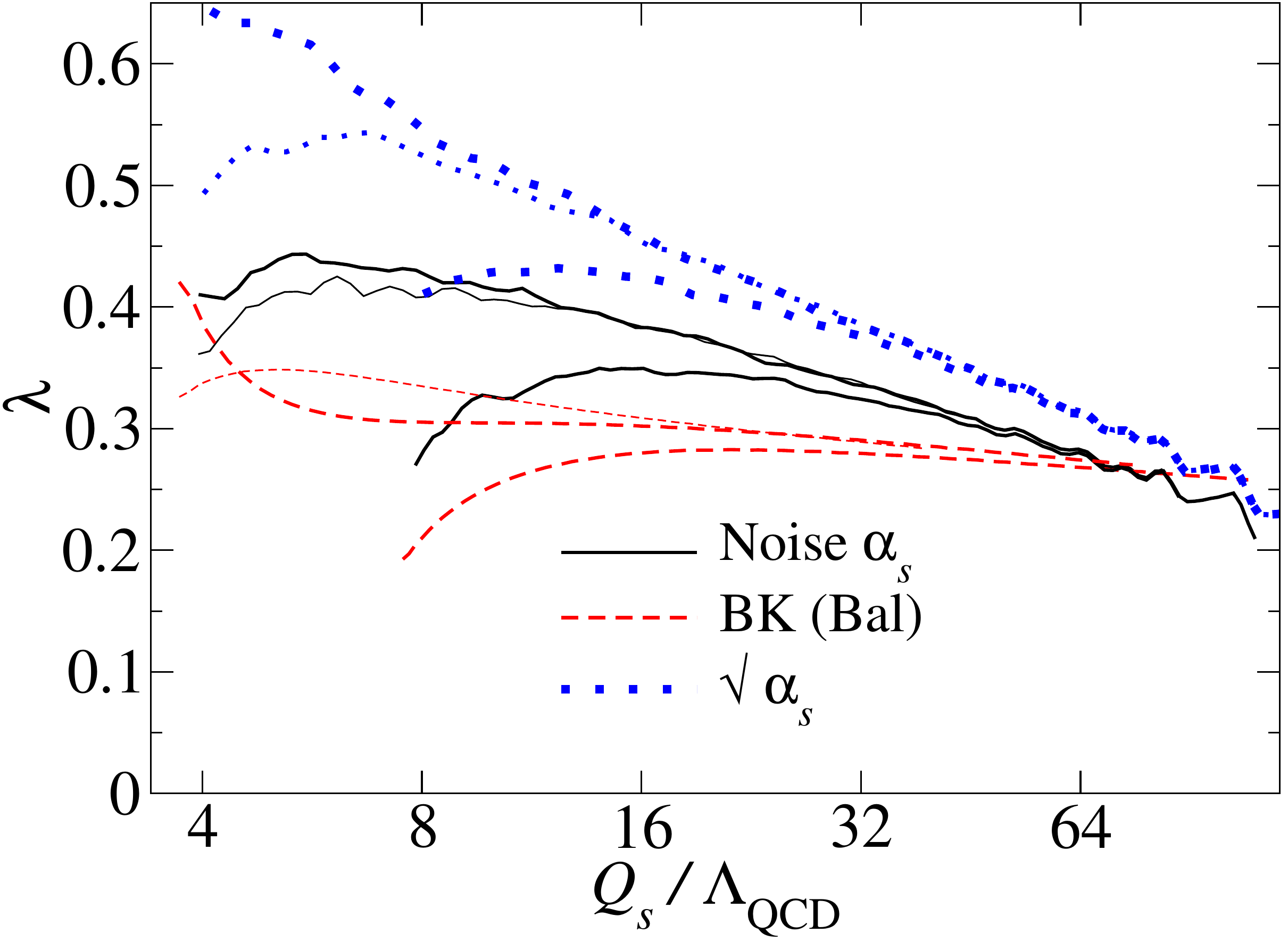}
\caption{\label{fig:lambda}
Evolution speed as defined in \eq\nr{eq:deflambda}. The thin line 
uses a different regularization of the Landau pole.
}
\end{wrapfigure}
Now $\xit_\ut$ corresponds to the emission of a gluon at
coordinate $\ut$ and the contraction with another $\xit_\vt$  
is interpreted as the absorption of the same gluon, 
the delta function in the $\xit\xit$-correlator is
a result of a  a sum over transverse momenta
of the gluon:
\begin{equation}
\delta^2(\ut-\vt) = \int \frac{\ud^2 \kt}{(2\pi)^2}
e^{i \kt\cdot(\ut-\vt)}.
\end{equation}
This interpretation breaks the explicit left-right symmetry
by choosing to measure the gluon momentum before or after
the interaction with the shockwave, i.e. on the left or right
of the Wilson line in \eq\nr{eq:jtimestepsymm}.
Figure~\ref{fig:virt} demonstrates the three possible two point 
contractions in between the $\xit$'s in the virtual contributions
(corrections to one Wilson line at $\xt$) and 
\fig\ref{fig:real} the corresponding real contributions (contractions
between $\xit$'s of two Wilson lines
at $\xt$ and $\yt$).

We will then rely on the general argument that the natural momentum scale 
for the running coupling is provided by the momentum of the emitted gluon.
This leads directly to our proposal for implementing the running
coupling in the JIWMLK equation, which is to regard the coupling constant as
a property of the correlator of the noise. We then propose to simply
replace $\xi$ by a new noise $\etat = ( \eta_1^at^a, \eta_2^a t^a)$ with
\begin{equation} \label{eq:newcorr}
\as \left\langle \xi_\xt^{a,i} \xi_\yt^{b,j} \right\rangle = \delta^{ab}
\as
\delta^{ij} \int \frac{\ud^2 \kt}{(2\pi)^2}
e^{i \kt\cdot(\xt-\yt)}
\longrightarrow  
\left\langle \eta_\xt^{a,i} \eta_\yt^{b,j} \right\rangle = \delta^{ab}
\delta^{ij} \int \frac{\ud^2 \kt}{(2\pi)^2}
e^{i \kt\cdot(\xt-\yt)} \as(\kt),
\end{equation}
in terms of which the rapidity step of the Wilson line is now
\begin{equation}\label{eq:jtimestepsymmeta}
V_{\xt}(y + \ud y) = 
\exp\left\{-i\frac{\sqrt{\ud y }}{\pi}\int_\ut 
  \Kt_{\xt-\ut} \cdot ( V_{\ut} \etat_{\ut} V^\dag_{\ut}) \right\}
V_{\xt}
\exp\left\{i\frac{\sqrt{\ud y }}{\pi}\int_\vt 
  \Kt_{\xt-\vt} \cdot \etat_{\vt} \right\}.
\end{equation}
% Note that since the noise correlator is not local in transverse coordinate,
% we have broken the left-right symmetry of the fixed 
% coupling equation. The physical interpretation of this
% is that the symmetry corresponds to time reversal, and using
% the momentum of a gluon in the final state after the target 
% (as opposed to an initial state one) 
% breaks the time reversal symmetry. While this loss of symmetry
% in inaesthetic, it seems inevitable at some point at least if
% one wants to generalize the evolution to more exclusive 
% observables (as discussed e.g. in \re\cite{Marquet:2010cf}).

Implementing the correlator
\nr{eq:newcorr} induces only a minor modification to the numerical
algorithm~\cite{Rummukainen:2003ns} used to solve the JIMWLK equation.
We denote the new noise correlator by
\begin{equation}\label{eq:newnoisecorr}
\astil_{\xt-\yt} \equiv \int \frac{\ud^2 \kt}{(2\pi)^2}
e^{i \kt\cdot(\xt-\yt)} \as(\kt),
\end{equation}
emphasizing  that $\astil_{\xt-\yt}$ is \emph{not}
the coupling evaluated at the scale $1/|\xt-\yt|$, but 
indeed the Fourier transform of $\as(\kt)$. Since $\as(\ktt)$ is a smooth, 
logarithmic function of $\ktt$, 
$\astil_{\xt-\yt}$ is very sharply peaked around $\xt=\yt$.

The proposed JIMWLK equation then results in the follosing evolution equation
for the dipole:
\begin{equation}\label{eq:bkfromj1}
\partial_y \hat{D}_{\xt-\yt}(y) 
=
\frac{\nc}{2\pi^2}
\int_{\ut,\vt} \astil_{\ut-\vt}
\Bigg\{
\left(
\Kt_{\xt-\ut}\cdot\Kt_{\xt-\vt} + \Kt_{\yt-\ut}\cdot\Kt_{\yt-\vt}
-2 \Kt_{\xt-\ut}\cdot\Kt_{\yt-\vt}
\right)
\frac{1}{2}\left[
\hat{D}_{\xt,\ut}\hat{D}_{\ut,\yt} + 
	\hat{D}_{\xt,\vt}\hat{D}_{\vt,\yt}
- \hat{D}_{\xt,\yt}
- \hat{D}_{\vt,\ut}\hat{Q}_{\xt,\vt,\ut,\yt}
\right],
\end{equation}
involving also the quadrupole 
$\hat{Q}_{\xt,\vt,\ut,\yt} = \frac{1}{\nc}\tr V^\dag_\xt V_\yt V^\dag_\ut V_\vt$.
 The integrand reduces to that of \eq\nr{eq:fixedas} in the limit $\ut=\vt$.

\section{Effect on the evolution}

We have shown in \re\cite{Lappi:2012vw} that parametrically the scale of the running 
coupling in \eq\nr{eq:bkfromj1} is the same as the ``Balitsky'' prescription, 
namely the size of smallest dipole in the problem.
Figure~\ref{fig:dipoles} shows the result of a numerical study of these different evolution 
equations. It is seen that the evolution is indeed slower than with the
previous ``square root'' coupling, and thus in better agreement with HERA data. It is, however,
not quite as slow as the BK equation with the ``Balitsky'' prescription.

The effect of the running coupling prescription on the evolution speed, parametrized
by
\begin{equation}\label{eq:deflambda}
 \lambda \equiv \frac{\ud \ln \qs^2}{\ud y}
\end{equation}
is shown in \fig\ref{fig:lambda}, where the above 
ordering is more explicit. Initially 
the evolution  speed depends on the initial condition, and 
for very small values of $\qs/\lqcd$ on the regularization of the Landau pole, 
while  at large values of $\qs$ lattice cutoff effects slow down 
the evolution in the JIMWLK case.

This work has been supported by the Academy of Finland, projects 
133005, 267321 and 273464 and by computing resources from
CSC -- IT Center for Science in Espoo, Finland. 

\bibliographystyle{h-physrev4mod2}
\bibliography{spires}






% \end{thebibliography}

\end{document}